\documentclass[12pt]{revtex4}

\usepackage{epsfig}

\begin{document}

\title{Comment on ``Quantitative x-ray photoelectron spectroscopy:
 Quadrupole effects, shake up, Shirley background, and relative
 sensitivity factors from a database of true x-ray photoelectron
 spectra"}

\author{
 M. B. Trzhaskovskaya $^1$ and V. K. Nikulin $^2$\\
\em $^1$ Petersburg Nuclear Physics Institute, Gatchina  188300, Russia
 \\
$^2$ Ioffe Physical Technical  Institute,  St.Petersburg 194021,
Russia}

\begin{abstract}
This Comment demonstrates that a comparison analysis by Seah and
Gilmore between experimental data on the X-ray photoelectron
spectroscopy intensities and theoretical data by Trzhaskovskaya
{\it et al.} is misleading due to a number of  serious errors made
by Seah and Gilmore (PRB 73 174113).
\end{abstract}
\maketitle

\hspace{9.cm} PACS number(s): 33.60.Fy

\vspace{0.5cm}

In a recent publication  by Seah and Gilmore \cite{seah}, a
comparison analysis is provided between  experimental X-ray
photoelectron spectroscopy  intensities measured at the National
Physical Laboratory and  theoretical data by Scofield
\cite{scof1,scof2} and by Trzhaskovskaya {\it et al.} \cite{t1,t2}.
Seah and Gilmore claim in the abstract that there is ``excellent
correlation between experimental intensities... and the
theoretical intensities involving the dipole approximation using
Scofield's cross sections. Here, more recent calculations
for cross sections  by Trzhaskovskaya {\it et al.} involving quadrupole
terms are evaluated and it is shown that their cross sections diverge
from the experimental database results by up to a factor of 5".

Another conclusion in \cite{seah} is concerned with  the photoionization
cross section $\sigma$ as well as the photoelectron angular
distribution parameters $\beta$ (the dipole parameter), $\gamma$, and
$\delta$ (quadrupole ones) obtained  by Scofield \cite{scof2} for Ne
and Ba at the photon energy $k=$ 3 keV: ``If these data are compared
with the data of Trzhaskovskaya {\it et al.} \cite{t1,t2}, good
agreement is found for $\beta$ and $\delta$, whereas $\gamma$ is
generally between 1.01 and 1.45 times greater and $\sigma$ is between
0.44 and 0.94 of these earlier values". Note that  here Seah and
Gilmore \cite{seah} have given  the wrong reference to one of
another Scofield's papers instead of \cite{scof2}.

We contend that the overall comparison  and  conclusions concerning
values of the photoionization cross section $\sigma$ and the
photoelectron angular distribution parameters $\beta, \gamma$,
and $\delta$ presented  in our papers \cite{t1,t2}, are invalid due to
serious errors and shortcomings made in \cite{seah}:

(i) Calculations by Scofield and by Trzhaskovskaya {\it et al.} are
compared in \cite{seah} for several values of the photon energy $k$, in
particular, for the $K_{\alpha}$ line of magnesium $k=$~1.254 keV and
for $k=$~3.0 keV. Photoionization cross sections \cite{scof1,scof2} and
the photoelectron angular distribution parameters \cite{scof2} are
presented by Scofield for these  values of the PHOTON ENERGY $k$.
In our papers \cite{t1,t2},  we give cross sections and angular
parameters for nine values of the PHOTOELECTRON KINETIC ENERGY
$E=k-\varepsilon_{b}$ where $\varepsilon_{b}$ is the binding
energy of the electron. This is pointed out everywhere in the text
of the papers from the title to the Section ``Explanation of Tables".
Nevertheless Seah and Gilmore \cite{seah} determine interpolated
values of $\sigma, \beta, \gamma$ and $\delta$ from  \cite{t1,t2} using
photoelectron energies $E$ as though they were photon energies $k$.

(ii) Comparing  the  photoionization cross
sections for an open atomic subshell, one should take into
consideration that values of $\sigma$   are given in
\cite{t1,t2,t3,band} for the completely filled subshells even though a
subshell is an open one. This is always pointed out in Section
``Explanation of Tables" of the papers. In contrast, Scofield
\cite{scof1} has not clearly indicated the manner in which  cross
sections   for the open relativistic doublet subshells have been
obtained. However analysis of the $\sigma$ values   from  \cite{scof1}
lead to the suggestion that he has calculated a combined
photoionization cross section $\sigma_{n\ell}$ per a real number of
electrons in two subshells with  total momenta $j_1=\ell-1/2$ and
$j_2=\ell+1/2$ where $n$ is the principal quantum number and $\ell$ is
the orbital momentum. Then $\sigma_{n\ell}$ has been spread between the
two relativistic subshells in accordance with their approximate
statistical weights.

Because of this, the only comparison of $\sigma_{n\ell}=\sigma_{n\ell
j_1}+\sigma_{n\ell j_2}$ for a specific number of electrons   is
meaningful.
This fact is disregarded in \cite{seah}. For the open subshell, Seah
and Gilmore compare $\sigma_{n\ell j}(S)$ of Scofield for a real
number of electrons as mentioned above, with $\sigma_{n\ell j}(T)$ given
in our tables \cite{t1,t2,band} for the completely filled subshell,
$\sigma_{n\ell j}(T)$ being found   mistakenly (see point (i)).

(iii) Besides, it is necessary to bear in mind  that different
theoretical assumptions may give rise to a difference in the results
obtained. In specific cases, the difference may be of great importance.
In the comparison of  results of the two calculations, Seah and Gilmore
do not consider the difference in models used by Scofield and  by
Trzhaskovskaya {\it et al.} \cite{t1,t2}.

Scofield has assumed the  electrons in the
initial and final  states treated as moving in the same central
Hartree-Dirac-Slater potential of the neutral atom (so-called the ``no
hole" model). By contrast,  we have taken into account the hole
in the atomic shell produced after ionization
(the ``hole" model) \cite{t1,t2,t3}. The hole has been
considered in the framework of the frozen orbital approximation \cite{t1}.
This is the only difference between
theoretical models used by Scofield and by Trzhaskovskaya {\it et al}.
Otherwise the atomic models are identical. In particular, both
calculations of  photoionization cross sections have been performed with
allowance made for all multipole orders of the photon field.

As to our calculations,
the subshell cross sections are calculated with a numerical accuracy
of 0.1$\%$. The  accuracy  has been verified \cite{rr} by comparing our
results  with benchmark relativistic calculations for one-electron
systems by Ichihara and Eichler \cite{ichihara}.

\vspace{-3.2cm}

\begin{figure}[h]
\centerline{
\epsfxsize=10cm\epsfbox{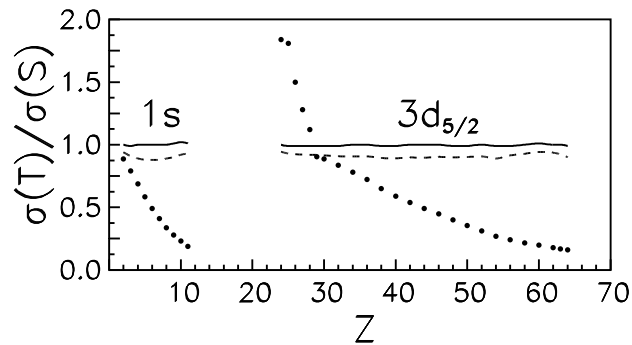}}
\vspace{-2.7cm}
\caption{\small Comparison of photoionization cross sections
obtained by Scofield \cite{scof1} $\sigma(S)$ without regard for the
hole in the atomic shell after ionization, with our values calculated
without $\sigma(T,nh)$ and with $\sigma(T,h)$ regard for the hole
 versus atomic number Z. The photon energy $k=$~1.254~keV. Solid,
$\sigma(T,nh)/\sigma(S)$; dashed, $\sigma(T,h)/\sigma(S)$; dark circles,
erroneous values presented by Seah and Gilmore in Fig.~4(a)
from \cite{seah}.}
\end{figure}

\vspace{-0.3cm}

In Fig.~1, the proper ratio of cross sections calculated by us  and by
Scofield \cite{scof1} $R_{\sigma}=\sigma(T)/\sigma(S)$  is presented
versus the atomic number $Z$
for the $1s$ and $3d_{5/2}$ shells at the photon energy $k=$~1.254~keV.
Our calculations have been performed in two different ways: using
exactly the same model as Scofield (see \cite{band,nef1}), that
is the ``no hole" model (solid lines) and using  the ``hole" model
\cite{t1,t2,t3} (dashed lines). Dark circles refer to the wrong ratio
$R_{\sigma}$ shown by Seah and Gilmore in Fig.~4(a) from \cite{seah}.
As evident from Fig.~1, solid lines practically coincide with the value
$R_{\sigma}=1.0$ because as has been shown earlier, our calculations
\cite{band,nef1} using the ``no hole" model agree with those by
Scofield \cite{scof1} within $\sim 1\%$. Dashed lines show that taking
the hole into account  results in a difference $\lesssim12\%$ in
values of $\sigma$ for the cases under consideration. Dark circles
located below $R_{\sigma}=1.0$ demonstrate that erroneous values of the
ratio presented by Seah and Gilmore diverge from correct values just by
up to a factor of 5. Dark circles located above $R_{\sigma}=1.0$
demonstrate the invalid comparison (see point (ii)) between $\sigma(T)$
and $\sigma(S)$ for the $3d_{5/2}$ subshell which is the open one for
elements with $Z\le28$.

We have also checked  that cross sections $\sigma$ for all
appropriate shells of Ne and Ba for the photon energy $k=3.0\,$keV
calculated by us using  the ``no hole" model agree with results by
Scofield \cite{scof2} also within $\sim 1\%$. A deviation of our values
of $\sigma $ obtained by the use of the ``hole" model from data
\cite{scof2} does not exceed $ 8\%$ rather than 56$\%$ claimed in
\cite{seah}.

\vspace{-3.2cm}

\begin{figure}[h]
\centerline{
\epsfxsize=10cm\epsfbox{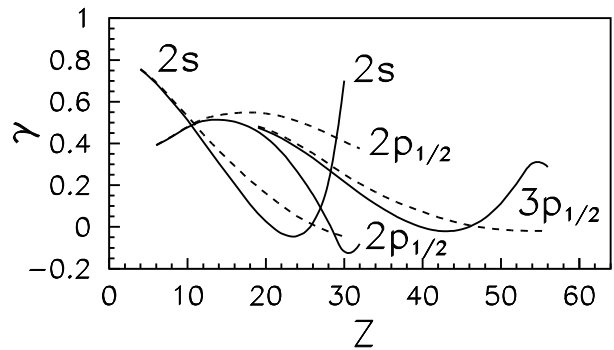}}
\vspace{-2.5cm}
\caption{\small
Non-dipole photoelectron angular distribution  parameter $\gamma$
 for the $2s, 2p_{1/2}$, and $3p_{1/2}$ shells versus atomic number $Z$.
Solid, calculations for the photon energy $k$=1.254~keV; dashed,
calculations for the photoelectron kinetic energy $E$=1.254~keV. }
\end{figure}

Values of photoelectron angular distribution parameters presented
in \cite{t1,t2} have also been  extracted by Seah and
Gilmore erroneously (see point (i)). We show in Fig.~2 our correct
calculations of parameter $\gamma$ (solid lines) and erroneous values
presented in Fig.~1(a) from \cite{seah} (dashed lines). The
$Z$-dependence of $\gamma$ is given for the photon energy $k$=1.254 keV
and for the $2s, 2p_{1/2}$, and $3p_{1/2}$ shells. As is seen in
Fig.~2, for a specific shell, solid and dashed lines coincide for low
$Z$ when the binding energy is small as compared with the photon
energy. As the binding energy increases  and the photoelectron
energy decreases, correct and erroneous curves become widely separated.
The maximum discrepancy may be much more than 1.45 which is pointed out
in \cite{seah}. Erroneous value of $\gamma$ may differ from correct
$\gamma$ up to many times and even change sign as in the
cases of the $2s$ and $2p_{1/2}$ shells.
It is obvious that the results presented in Figs.~1(b), 4(b), 4(c),
5, 11, and 12 of paper \cite{seah} are erroneous for the same reason.

 Comparison between our calculations of the non-dipole
parameters $\gamma$ and $\delta$ and calculations by Scofield
\cite{scof2} for several subshells of neon, copper, and barium is
presented in Table~I.

\begin{table}
\caption{
 Ratios of the non-dipole photoelectron angular
distribution parameters $\gamma$ and $\delta$ calculated by us (T) with
and without regard for the hole to those calculated by
Scofield (S) without regard for the hole \cite{scof2}. Photon energy
$k$=3.0 keV. $R_{\gamma}=\gamma(T)/\gamma(S)$ and
$R_{\delta}=\delta(T)/\delta(S)$.}

\begin{center}
\begin{tabular}{clcccc} \hline\hline
  &  & \multicolumn{2}{c}{no hole}& \multicolumn{2}{c}{hole}\\
\cline{3-6}
 Z & Shell & $R_{\gamma}$ & $R_{\delta}$ & $R_{\gamma}$
 & $R_{\delta}$\\
 \hline
10 &   $1s$ & 1.00 &  & 1.00 &  \\
 &   $2s$ & 1.00 &  & 1.00 & \\
 &   $2p_{1/2}$ & 1.04 &0.94  & 1.03 & 0.94\\
\smallskip
 &   $2p_{3/2}$ & 1.03 &0.96  & 1.02 & 0.96\\
 29&   $2s$ & 0.98 &  & 1.44 & \\
 &   $2p_{1/2}$ & 1.01 &0.96  & 1.01 & 0.95\\
 &   $2p_{3/2}$ & 1.00 &0.96  & 1.01 & 0.97\\
 &   $3s$ & 0.99 &  & 1.02 & \\
 &   $3p_{1/2}$ & 1.01 &0.94  & 1.00 & 0.95\\
 &   $3p_{3/2}$ & 1.00 &0.94  & 1.00 & 0.95\\
 &   $3d_{3/2}$ & 1.03 &0.94  & 1.03 & 0.94\\
\smallskip
 &   $3d_{5/2}$ & 1.02 &0.95  & 1.02 & 0.95\\
56 &   $3s$ & 1.03 &  & 0.77 & \\
 &   $3p_{3/2}$ & 1.20 &0.90  & 1.98 & 0.95\\
 &   $3d_{3/2}$ & 1.01 &0.96  & 1.02 & 0.95\\
 &   $3d_{5/2}$ & 1.00 &0.98  & 1.00 & 0.98\\
 &   $4s$ & 1.00 &  & 1.02 & \\
 &   $4p_{1/2}$ & 1.04 &  & 1.10 & \\
 &   $4p_{3/2}$ & 1.05 &1.09  & 1.10 & 1.09\\
 &   $4d_{3/2}$ & 1.00 &0.97  & 1.01 & 0.97\\
 &   $4d_{5/2}$ & 1.00 &0.98  & 1.00 & 0.98\\
 &   $5s$ & 0.97 &  & 0.97 & \\
 &   $5p_{1/2}$ & 1.05 &  & 1.06 & \\
 &   $5p_{3/2}$ & 1.02 &0.94  & 1.03 & 0.94\\
\hline\hline
\end{tabular}
\end{center}
\end{table}

 We list ratios
$R_{\gamma}=\gamma(T)/\gamma(S)$ and
$R_{\delta}=\delta(T)/\delta(S)$ where our calculations (T)
have been performed using two models: ``no hole" and "hole". We omit
cases where   magnitudes of $\gamma$ and  $\delta$ are very close to
zero. Photon energy is equal 3.0 keV.

Table I demonstrates that there is good agreement between calculations
\cite{scof2} and our results obtained in the same ``no hole" model.
As is seen, in the majority of
cases, the parameters $\gamma$ and $\delta$ are little affected by
taking the hole into account. However there exists a considerable
deviation of $R_{\gamma}$ from the value $R_{\gamma}=1.0$ for several
cases, for example, $R_{\gamma}$=1.98 for the $3p_{3/2}$ shell in Ba,
$R_{\gamma}$=1.44 for the $2s$ shell in Cu, and
$R_{\gamma}$=0.77 for the $3s$ shell in Ba.

\vspace{-2.7cm}

\begin{figure}[h]
\centerline{
\epsfxsize=10cm\epsfbox{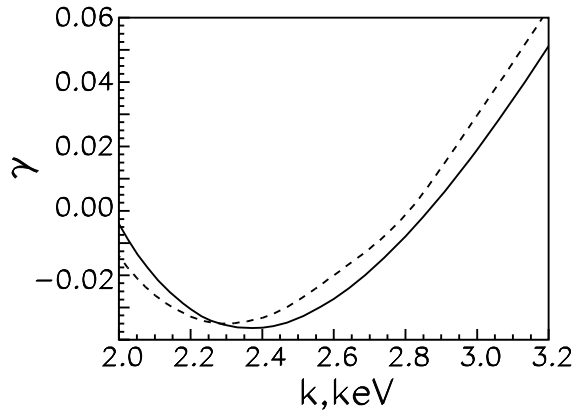}}
\vspace{-2.1cm}
\caption{\small
The $k$-dependence of the parameter $\gamma$ for the
$3p_{3/2}$ shell of the barium atom. Solid, calculations without
regard for the hole after ionization; dashed, with regard for the hole.
}
\end{figure}

This difference is associated with the fact that the $k$-dependence of
$\gamma$ has a
minimum not far from $k$=3.0~keV. So curves $\gamma(k)$ obtained with
and without regard for the hole are shifted relative to each other as
is seen in Fig.~3 for the  $3p_{3/2}$ shell of barium.
Nonmonotonous behaviour of the parameters when curves $\beta(k),
\gamma(k)$, and $\delta(k)$ may take the form of oscillations has been
discussed at length in our paper \cite{jphysb}. In such cases, all
assumptions underlying the calculation, a minor difference in binding
energies, and other calculational details may have a great impact on
values of the parameters.

In summary it should be emphasized that we have clearly
demonstrated  that dramatic deviations of the results by
Trzhaskovskaya {\it et al.} from those by Scofield  and consequently,
the deviation of our results from  experimental data reported by Seah
and Gilmore, do not actually take place and are due to  errors
(i) and (ii) and shortcomings (iii) in paper \cite{seah}. Reasonable
deviations between the two calculations are associated with somewhat
different atomic models used in \cite{scof1,scof2} and in
\cite{t1,t2,t3}.

\newpage
This work is partially supported by  International Atomic Energy Agency
(Contract No.\,13349/RBF) and  Russian Foundation  for Basic Research
(project No.\,06-02-16489).

\end{document}